\documentclass[
superscriptaddress,
amsmath, amssymb,
aps, physrev,
prb,
twocolumn,
]{revtex4-2}

\usepackage{graphicx}
\usepackage{dcolumn}

\usepackage{algorithm}
\usepackage{algpseudocode}

\graphicspath{{figures/}}
\usepackage[separate-uncertainty=false]{siunitx}

\usepackage[capitalise]{cleveref}

\usepackage{changes}
\definechangesauthor[name={Dimitris}, color=orange]{dt}

\newcommand{\basis}{cc-pVDZ }
\DeclareSIUnit\angstrom{\text{Å}}

\begin{document}


\title{Enhancing quantum-classical configuration interaction methods using a neural-network classifier}

\author{Severino Zeni}
\email[]{severino.zeni@exact-lab.it}
\affiliation{eXact lab Srl, Via Francesco Crispi 56,  34126 Trieste, Italy}

\author{Giovanni Varutti}
\affiliation{eXact lab Srl, Via Francesco Crispi 56,  34126 Trieste, Italy}
\affiliation{Department of Physics, University of Trieste, Strada Costiera 11, 34151 Trieste, Italy}

\author{Jacopo Nespolo}
\affiliation{eXact lab Srl, Via Francesco Crispi 56,  34126 Trieste, Italy}

\author{Dimitrios Trypogeorgos}
\affiliation{CNR Nanotec, Institute of Nanotechnology, via Monteroni, 73100, Lecce, Italy}

\date{\today}

\begin{abstract}
Selected configuration interaction methods achieve near-exact electronic structure calculations by iteratively constructing compact variational spaces, but their efficiency depends critically on the heuristics used to identify important determinants. 
Here, we introduce a data-driven selection framework that recasts determinant importance as a binary classification task and integrates a neural-network classifier into the iterative CI workflow through an active-learning loop. 
At each iteration, a random subset of candidate determinants is labelled via temporary diagonalisation, and the trained classifier guides selection of the remaining configurations.
We demonstrate the utility of this framework for both classical and quantum CI methods by calculating the ground-state energy of a diatomic molecule.
Our method achieves result parity with traditional configuration interaction methods at substantially lower computational cost: roughly a $\times 5$ reduction in memory and per-iteration cost for the classical cHCI variant, and convergence in markedly fewer iterations for the quantum-classical cSQD variant.
These results establish classifier-assisted determinant selection as a lightweight, method-agnostic tool for compressing variational spaces and accelerating both classical and hybrid quantum-classical configuration interaction algorithms.
\end{abstract}


\maketitle

\section{Introduction}\label{sec:intro}

Accurate electronic-structure calculations require methods capable of describing electron correlation within Hilbert spaces whose dimension grows exponentially with the number of active orbitals. 
While full configuration interaction (FCI) provides an exact treatment within a fixed one-particle basis, its computational cost limits its use to very small systems. 
Selected configuration interaction (SCI) methods offer an attractive compromise by constructing compact variational spaces containing only the most important determinants~\cite{ivanic_identification_2001}. 
Through iterative selection guided by perturbative criteria or excitation heuristics, cf. Fig. \ref{fig:1}(a), SCI methods such as CIPSI~\cite{huron_iterative_1973}, ASCI~\cite{tubman_deterministic_2016}, and HCI~\cite{holmes_heat-bath_2016} can achieve near-FCI accuracy at a fraction of the cost. 

Nevertheless, SCI calculations must repeatedly evaluate or screen large pools of candidate determinants, and the quality of the result depends critically on the efficiency and accuracy of the selection mechanism.
As the selected space grows, the number of connected determinants that must be inspected at each iteration can become large, which motivates several refinements that have appeared in the literature. 
For example, semi-stochastic variants of HCI~\cite{sharma_semistochastic_2017} introduce stochastic sampling only in the perturbative stage to alleviate memory requirements, while retaining the deterministic variational selection. 
Other extensions have explored strategies for accelerating excitation generation or improving thresholding schemes without altering the basic selection logic~\cite{li_fast_2018}. 
Despite such developments, the core variational mechanism still relies on enumerating and evaluating all candidate determinants above the screening threshold. 

Quantum computers could offer a valid solution to the scaling problem since they promise to directly solve the electronic Schr\"odinger equation in exponentially large Hilbert spaces.
Quantum SCI methods, usually referred to as sample-based quantum diagonalisation (SQD)~\cite{robledo-moreno_chemistry_2025}, have emerged as promising near-term approaches to solving the CI problem using NISQ devices.
The SQD approach approximates the ground state of an electronic Hamiltonian $H$ by constructing a subspace from measurements performed on a quantum device and diagonalising the Hamiltonian projected onto that subspace.
SQD is generally inefficient in finding new determinants as sampling repeatedly selects the same configurations. 
This repetition becomes more pronounced when sampling from an approximate ansatz and ultimately limits the accuracy of the calculations~\cite{reinholdt_fundamental_2025}.

This creates an opportunity for methods that can guide or augment the selection step directly, potentially reducing the number of determinants that must be examined while preserving the structure of a typical CI iteration. 
The classifier-based modification introduced here follows this direction by providing an alternative means of identifying promising determinants during the variational-space expansion.

\section{A neural-network framework}

Recent developments in machine learning offer new opportunities for accelerating quantum chemistry methods through improved sampling, prediction, and compression. 
In particular, neural networks have shown promise as flexible function approximators capable of learning from the structure of electronic wave functions, aiding in tasks such as determinant selection, excitation ranking, and orbital optimisation. 
A growing body of work has begun exploring neural-network–based configuration classifiers that operate directly on occupation-number bitstrings, providing a data-driven alternative to traditional perturbative or energy-based selection criteria
~\cite{coe_machine_2018, coe_machine_2019, jeong_active_2021, pineda_flores_chembot_2021, goings_reinforcement_2021, herzog_solving_2023, bilous_deep-learning_2023, bilous_neural-network_2024, bilous_neural-network-supported_2025, schmerwitz_neural-network-based_2025-1}. 
Bitstring-based neural classifiers, in particular, provide a flexible, architecture-agnostic way to operate directly on determinantal occupations. 
These developments suggest that data-driven selection may complement traditional SCI heuristics and help construct compact, accurate wave functions more efficiently.

Here, we adopt a unified view of determinant selection as a binary classification problem and introduce a general active-learning framework that integrates naturally with iterative classical and quantum SCI schemes, depicted in Fig. \ref{fig:1}(b). 
At each iteration, the underlying method generates a pool of candidate determinants. 
A small random subset is added to the current variational space, a temporary diagonalisation yields CI coefficients for this subset, and these coefficients are converted into balanced importance labels. 
A neural-network classifier is trained from scratch on this labelled data and then applied to the remaining candidates to predict which ones should be included in the next variational update. 
This procedure focuses computational effort on a small labelled subset while using the classifier to guide selection across the full pool. 
Because training and inference occur within each iteration, the classifier continually adapts to the evolving structure of the approximate ground-state wave function.

We apply this framework in two distinct families of SCI-inspired quantum chemistry algorithms: a classifier-enhanced variant of the heat-bath configuration interaction method (cHCI) and a classifier-enhanced variant of sample-based quantum diagonalisation (cSQD).

We use our classifier-enhanced methods to calculate the ground state energy of the N$_2$ molecule and benchmark against traditional methods.
The cHCI method recovers more than 98\% of the CASCI correlation energy using only $1/5$ of the determinants required by a converged HCI calculation, yielding a comparable $\sim 5\times$ reduction in both the memory footprint and the per-iteration cost of the variational Davidson diagonalisation.
Similarly, cSQD achieves substantially faster convergence than standard SQD, reaching below 1\% relative correlation-energy error in just 3 iterations---a threshold that standard SQD fails to cross even after 20 iterations.

\section{Slater determinant classification}\label{classifier}

Machine-learning classification methods provide a flexible way to identify important configurations in configuration interaction calculations. 
Because only a small fraction of all determinants contribute appreciably to the ground-state wave function~\cite{ivanic_identification_2001}, the task can be framed as a binary classification problem in which the aim is to distinguish important configurations from the overwhelmingly more numerous unimportant ones. 

\begin{figure}[htb]
	\begin{center}
	   \includegraphics[width=1\columnwidth]{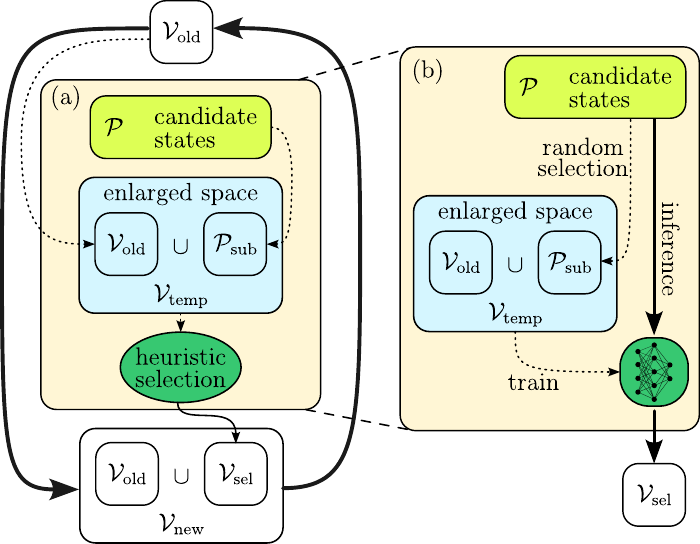}
	\caption[]{Generic iteration of selected CI schemes. (a) Standard algorithms enlarge the current variational space $\mathcal{V}_\mathrm{old}$ with a subset $\mathcal{P}_\mathrm{sub}$ of the candidate pool $\mathcal{P}$ (the construction of both is scheme-dependent). A heuristic then selects from the enlarged space $\mathcal{V}_\mathrm{temp}$ the states $\mathcal{V}_\mathrm{sel}$ that enter the updated space $\mathcal{V}_\mathrm{new}$, which seeds the next iteration. (b) Our framework replaces the selection step: a random subset $\mathcal{P}_\mathrm{sub}$ ($\mathcal{P}_\mathrm{rand}$ in the text) is drawn from $\mathcal{P}$, a single diagonalisation in $\mathcal{V}_\mathrm{temp}$ assigns importance labels to its states, and the labelled data are used to train a neural-network classifier. The trained network then scores the remaining candidates in $\mathcal{P}$, and the highest-ranked states form $\mathcal{V}_\mathrm{sel}$.}
	\label{fig:1}
	\end{center}
\end{figure}

In the general framework adopted here, each configuration is mapped to a numerical representation suitable for neural-network processing, and the classifier predicts whether the configuration is likely to carry significant weight in the wave function. 
The classifier is not asked to predict CI coefficients directly; instead, it outputs a probability or score indicating whether a configuration should be included in the variational space of the next iteration. 
The key ingredient is the availability of labels that reflect the actual importance of configurations at the current stage of the calculation. 
These labels are generated from a small diagonalisation in a temporarily expanded space, providing a consistent and method-agnostic source of training data.

The classifier enters the algorithm through an active-learning loop, see \cref{fig:1}(b).
Each iteration begins with a current variational space $\mathcal{V}_{\mathrm{old}}$ and a pool $\mathcal{P}$ of candidate configurations generated by the underlying algorithm, for example, excitations identified by heat-bath screening in HCI or determinants produced by configuration recovery in SQD.
The construction of $\mathcal{P}$ depends on the specific method under consideration, but in all cases it represents the set of configurations from which the variational space may be extended at the current iteration.

Rather than immediately expanding the variational space to include the full pool $\mathcal{P}$, a temporary training space is constructed by selecting a fixed fraction $f_\mathrm{s}$ of configurations from $\mathcal{P}$ at random. 
Denoting this subset by $\mathcal{P}_{\mathrm{rand}}$, the temporary variational space is defined as
\begin{equation}
\mathcal{V}_{\mathrm{temp}} = \mathcal{V}_{\mathrm{old}} \cup \mathcal{P}_{\mathrm{rand}} .
\end{equation}
The Hamiltonian projected onto $\mathcal{V}_{\mathrm{temp}}$ is diagonalised once, and the resulting expansion coefficients are used to assign importance labels to the configurations in $\mathcal{P}_{\mathrm{rand}}$. 
Configurations whose amplitudes exceed a chosen threshold are labelled as important, while the remaining ones are labelled as unimportant. 
The threshold is selected so as to produce a balanced set of important and unimportant configurations, which is advantageous for training the classifier.

A neural-network classifier is then trained from scratch at the current iteration using the labelled configurations in $\mathcal{P}_{\mathrm{rand}}$. 
After training, the classifier is applied to the remaining configurations in the pool,
\begin{equation}
\mathcal{P}_{\mathrm{rest}} = \mathcal{P} \setminus \mathcal{P}_{\mathrm{rand}},
\end{equation}
yielding a predicted probability of importance for each configuration.

The variational object used by the next iteration is constructed by combining two contributions.
First, all configurations in $\mathcal{P}_{\mathrm{rand}}$ that were labelled as important during the temporary diagonalisation are retained.
Second, a method-dependent number $M$ of configurations from $\mathcal{P}_{\mathrm{rest}}$ with the highest classifier scores is added.
The updated space is therefore given by
\begin{align}
\mathcal{V}_{\mathrm{new}} = \mathcal{V}_{\mathrm{old}}
&\cup \left\{ D \in \mathcal{P}_{\mathrm{rand}} \;:\; D \text{ labelled important} \right\} \nonumber \\
&\cup \left\{ D \in \mathcal{P}_{\mathrm{rest}} \;:\; \text{top-}M \text{ predictions} \right\}.
\end{align}
The algorithm then proceeds to the next iteration by diagonalising the Hamiltonian in $\mathcal{V}_{\mathrm{new}}$.
The two methods we consider differ in what $\mathcal{V}_{\mathrm{old}}$ denotes and in how $M$ is chosen: in cHCI, $\mathcal{V}_{\mathrm{old}}$ is the cumulative variational space and $M = |\mathcal{P}_{\mathrm{rand}}|$, so the space grows by $|\mathcal{P}_{\mathrm{rand}}|$ classifier-selected determinants per iteration; in cSQD, $\mathcal{V}_{\mathrm{old}}$ is the batch used in the previous iteration and $M = |\mathcal{P}_{\mathrm{rand}}| - N_{\mathrm{imp}}$, so the new batch retains the size $|\mathcal{P}_{\mathrm{rand}}|$ regardless of the number $N_{\mathrm{imp}}$ of important configurations identified in $\mathcal{P}_{\mathrm{rand}}$.
The two instantiations are detailed in Secs.~\ref{sec:chci} and~\ref{sec:csqd}.

This procedure defines a general active-learning loop for configuration selection: the randomly selected subset $\mathcal{P}_{\mathrm{rand}}$ supplies explicit labels from a single diagonalisation, while the classifier --- retrained from scratch at each iteration so that it tracks the evolving wave function --- guides the selection of additional configurations from the candidate pool.
Because the framework modifies only the selection step and the classifier is inexpensive to evaluate, it can be combined with a wide range of classical and quantum-assisted iterative diagonalisation methods without altering their core workflow.

\section{Numerical results}

\begin{figure*}[htb]
	\begin{center}
	 \includegraphics[width=2\columnwidth]{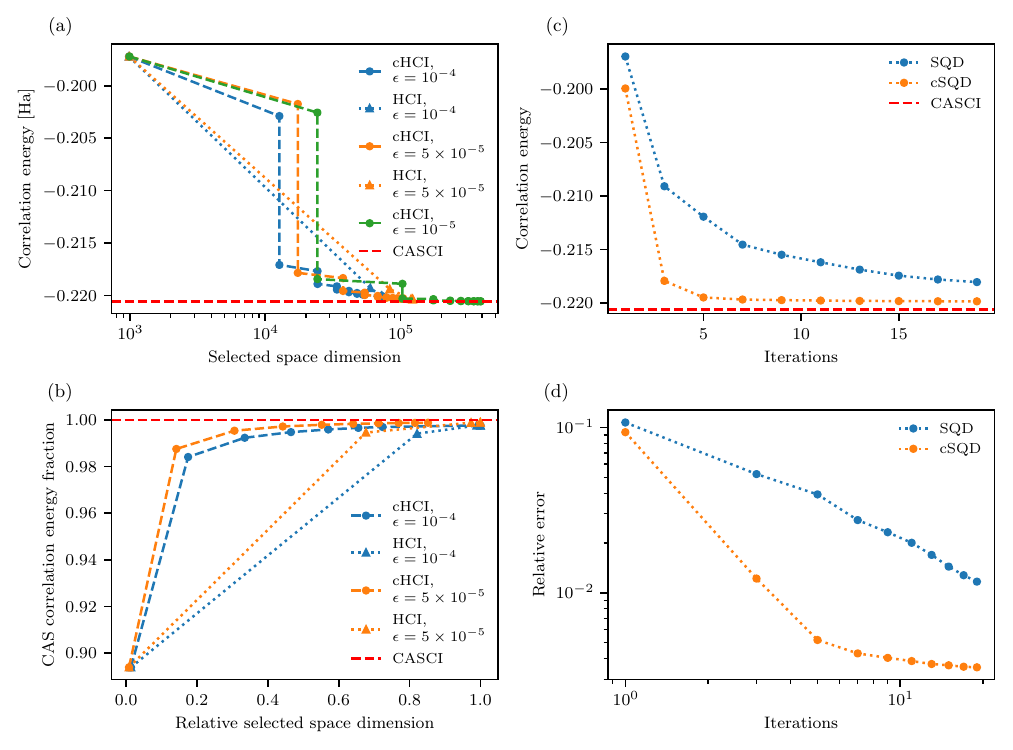}
	\caption[]{(a) N$_2$ ground state correlation energy as a function of the selected space size.
    The cHCI curves exhibit a stepped behaviour that is due to the successive steps of random sampling of the HCI pool and inferred selection, with the vertical drops representing the gains in correlation energy provided by the NN classifier.
    (b) N$_2$ ground state correlation energy (as a fraction of the CASCI value) as a function of the space dimension relative to the converged HCI space dimension.
    The plot highlights that cHCI can recover $> 98 \%$ of the correlation energy with a space dimension that is under $1 / 5$ of that of the converged HCI calculation.
    (c) N$_2$ ground state correlation energy as a function of the selected space size.
    (d) N$_2$ ground state correlation energy relative error (compared to the CASCI value) as a function of the iteration number. The cSQD data show improved convergence.
    All cHCI panels use a selection fraction $f_\mathrm{s} = 0.2$, and all SQD/cSQD panels use a carryover threshold of $10^{-4}$.
    The data for all plots were computed at the equilibrium internuclear distance $R = \qty{2.1}{\angstrom}$, in the \basis basis, with $N_\mathrm{o} = 16$ active spatial orbitals and $N_\mathrm{e} = 10$ active electrons.}
	\label{fig:2}
	\end{center}
\end{figure*}

All one- and two-electron integrals used in this work are obtained from PySCF Hartree--Fock calculations~\cite{sun_pyscf_2018}. 
Standard HCI reference results, as well as the excitation generation and variational diagonalisations employed in the cHCI calculations, are performed using the \texttt{naive-hci} extension of PySCF, which solves the variational eigenproblem with the Davidson algorithm.
Standard SQD calculations use the \texttt{qiskit-addon-sqd} package~\cite{qiskit-addon-sqd}; cSQD reuses its sampling and configuration-recovery routines and replaces the batch-construction step with the classifier-driven procedure described above.
All cHCI and cSQD results below use a random-subset fraction $f_\mathrm{s} = 0.2$.

\subsection{cHCI}\label{sec:chci}

The cHCI method modifies only the variational stage of standard HCI. 
Each iteration begins by generating a pool of candidate configurations.
This procedure collects all single and double excitations whose off-diagonal Hamiltonian elements satisfy the HCI screening criterion. 
The resulting set forms the pool $\mathcal{P}$ of determinants that standard HCI would examine for inclusion through the importance measure $|H_{ki} c_i|$.
Following the procedure of Sec.~\ref{classifier}, the variational space is then expanded by retaining all configurations in $\mathcal{P}_{\mathrm{rand}}$ that were labelled important and adding the $M = |\mathcal{P}_{\mathrm{rand}}|$ highest-scoring configurations from $\mathcal{P}_{\mathrm{rest}}$, so the space grows by $|\mathcal{P}_{\mathrm{rand}}|$ classifier-selected determinants per iteration.
In early iterations where the pool is small (specifically, $|\mathcal{P}| < 20{,}000$), we skip the classifier step and include the entire pool directly in the variational space.

We benchmark cHCI against standard HCI on the ground state of the N$_2$ molecule at the equilibrium internuclear distance $R = \qty{2.1}{\angstrom}$. 
For several values of the HCI selection parameter $\epsilon_1$, we compare:
\begin{itemize}
    \item the correlation energy recovered by cHCI versus HCI as a function of variational-space dimension in \cref{fig:2}(a), and
    \item the correlation-energy error relative to CASCI correlation energy in \cref{fig:2}(b).
\end{itemize}
Across all values of $\epsilon_1$ studied, cHCI achieves substantial wave function compression. 
In particular, in the case reported here, cHCI recovers more than 98\% of the CASCI correlation energy using a variational space that is only $1/5$ of the size of the converged HCI variational space. 
This demonstrates that the classifier-enhanced selection mechanism retains the essential determinants required for an accurate description of the ground-state wave function while significantly reducing the cost of the variational-stage diagonalisations.

\subsection{cSQD}\label{sec:csqd}

We introduce a classifier-driven approach to guiding the construction of the SQD subspaces. It can be used as an alternative to the standard SQD batching procedure --- as in our single-batch implementation here --- or in conjunction with it, with each batch assembled by the classifier-driven procedure described below. 
Since diagonalisation is the dominant computational cost in SQD and scales with the dimension of the subspace on which the Hamiltonian is projected, it is desirable to restrict attention to determinants that are most likely to influence the ground-state wave function.

The cSQD method modifies only the stage of the SQD workflow in which determinantal subspaces are constructed for the projected Hamiltonian diagonalisation. 
All other components of SQD --- quantum sampling, configuration recovery, and occupation updating --- follow the standard method described in~\cite{robledo-moreno_chemistry_2025}. 
As in SQD, each iteration begins with a recovered ensemble $X_R$ of symmetry-corrected configurations obtained from the noisy quantum samples. 
This ensemble serves as the pool of candidate determinants, denoted $\mathcal{P}$.
The batch used in the previous iteration plays the role of $\mathcal{V}_{\mathrm{old}}$ in the temporary training space $\mathcal{V}_{\mathrm{temp}} = \mathcal{V}_{\mathrm{old}} \cup \mathcal{P}_{\mathrm{rand}}$ of Sec.~\ref{classifier}.

The batch used for the projected Hamiltonian diagonalisation is then assembled following the procedure of Sec.~\ref{classifier}: it retains all $N_{\mathrm{imp}}$ configurations from $\mathcal{P}_{\mathrm{rand}}$ that were labelled important and is topped up with the $M = |\mathcal{P}_{\mathrm{rand}}| - N_{\mathrm{imp}}$ highest-scoring configurations from $\mathcal{P}_{\mathrm{rest}}$, so the new batch retains the size $|\mathcal{P}_{\mathrm{rand}}|$ regardless of $N_{\mathrm{imp}}$.
Because our implementation uses a single batch, the orbital-occupation numbers entering the next configuration-recovery step are computed from the eigenvector obtained in this batch alone.

Although SQD is designed for use with bitstrings obtained from quantum measurements, the benchmarks reported here use bitstrings drawn from a uniform classical distribution. 
This allows us to isolate the algorithmic behaviour of SQD and cSQD independently of hardware noise~\footnote{In a separate study~\cite{Varutti_in_preparation}, we show that, with currently available circuits and sampling techniques, we do not observe advantages from quantum-hardware sampling relative to this classical baseline.}.

We benchmark cSQD against standard SQD on the ground state of the N$_2$ molecule at the equilibrium internuclear distance $R = \qty{2.1}{\angstrom}$. 
For both methods, we track the correlation-energy error relative to the CASCI correlation energy as a function of the SQD iteration number. 
The results show that the classifier-enhanced construction of the batch subspace accelerates convergence dramatically (\cref{fig:2}(c,d)).
In the configuration studied here, cSQD reaches a relative correlation-energy error below 1\% after only 3 iterations, whereas standard SQD does not achieve this accuracy even after 20 iterations.

\section{Conclusions and outlook}

Machine-learning classifiers have recently emerged as a promising tool for assisting determinant selection in selected configuration interaction and related iterative methods. 
Here, we have explored how a neural-network classifier can be integrated into two established workflows, HCI and SQD, through a shared active-learning procedure that relies only on a small, randomly chosen subset of determinants at each iteration. 
The resulting variants, cHCI and cSQD, retain the structure of their parent algorithms while modifying only the construction of the variational space.

We demonstrated the effectiveness of this approach by calculating the ground state of the N$_2$ molecule.
Our results illustrate how neural-network classifiers can serve as powerful tools for guiding determinant selection and recovery in SCI algorithms. 
In cHCI, classifier-guided selection produces a substantial reduction of the variational-space dimension while maintaining high accuracy: the method recovers more than 98\% of the CASCI correlation energy using a space $1/5$ of the size required by a converged HCI calculation. 
In cSQD, classifier-based selection of determinants for the projected diagonalisation yields markedly faster convergence: the method reaches below 1\% relative error in 3 iterations, while standard SQD fails to reach this threshold within 20 iterations.
Beyond the specific applications explored here, these results suggest a broader role for machine learning in designing compact wave function representations and more efficient electronic structure workflows.

At the same time, our neural-network framework represents an initial step toward a more comprehensive understanding of classifier-assisted determinant selection. 
The numerical calculations shown here focus on a single molecular system and on the variational stage of the algorithms; it would be interesting to assess performance across a broader range of molecules, active spaces, and correlation regimes, and to investigate how the approach interacts with perturbative corrections or alternative labelling strategies.
We expect that further development and broader testing of this framework will contribute to scalable and data-informed approaches to electronic-structure calculations.

\begin{acknowledgments}
We acknowledge fruitful discussions with Simone De Liberato and Layla Martin-Samos. 
S.Z. thanks Chiara Arduini for assistance in creating \cref{fig:1}.
Funding was provided by the ICSC---Centro Nazionale di Ricerca in High Performance Computing, Big Data and Quantum Computing, funded by European Union---NextGenerationEU-PNRR, Missione 4 Componente 2 Investimento 1.4 Grant number CN00000013 via project JANAS-QMLMS (CUP:B93C22000620006, COR:23037841); 
the Italian Ministry of University and Research (MUR) under the granting scheme FIS 3 (grant number FIS-2024-04047).
\end{acknowledgments}

\section*{Data availability}
The data and code that support the findings of this article are available
from the corresponding author upon reasonable request.

\clearpage
\newpage

\appendix
\section{Heat-bath configuration interaction \label{sec:hci}}

HCI~\cite{holmes_heat-bath_2016} is a selected configuration interaction plus perturbation theory method that seeks to approximate the FCI solution by constructing, in a controlled manner, a reduced variational space of important determinants. 
As in other SCI schemes, the variational stage is used to build an approximate ground-state wave function through an iterative expansion of the determinant space, while a separate perturbative stage refines the energy. 
In contrast to earlier approaches such as CIPSI~\cite{huron_iterative_1973} and ASCI~\cite{tubman_deterministic_2016}, which evaluate perturbative amplitudes for all excitations connected to the current wave function, HCI introduces a simple screening principle that restricts attention to excitations associated with sufficiently large Hamiltonian matrix elements. 
This screening allows the algorithm to target the dominant contributors to the variational wave function without examining the full excitation structure of the Hamiltonian.

The variational stage of HCI proceeds by iteratively diagonalising the Hamiltonian within the current selected space and then augmenting this space according to a prescribed importance criterion. 
After each diagonalisation, the coefficients ${c_i}$ of the current variational eigenvector are used to identify new determinants $D_k$ through the measure $\lvert H_{ki}\, c_i \rvert$, where $H_{ki}$ is the Hamiltonian matrix element between the $k$-th and $i$-th Slater determinants $H_{ki} = \langle D_k \rvert H \lvert D_i \rangle $.
A determinant is added whenever
\begin{equation}
\lvert H_{ki}\, c_i \rvert > \epsilon_1
\end{equation}
for at least one determinant $D_i$ already in the space. 
The parameter $\epsilon_1$ thus controls the granularity of the variational expansion: large values produce a smaller, coarser selected space, while decreasing $\epsilon_1$ systematically refines the space and, in the limit $\epsilon_1 \to 0$, the procedure recovers the FCI space. 
The diagonalise–select cycle is repeated until a chosen convergence criterion is met, at which point the variational wave function is taken as the result of the selection process.
While the variational stage of HCI provides a systematic route toward the FCI limit, its performance remains tied to the deterministic selection criterion and to the explicit generation of all excitations that satisfy $\lvert H_{ki}\, c_i \rvert > \epsilon_1$.

\section{Sample-based quantum diagonalisation}

A parameterised quantum circuit prepares a state $|\tilde{\Psi}\rangle$ intended to approximate the ground-state wave function, and repeated measurements yield bitstrings $x$ drawn from the distribution $P_{\tilde{\Psi}}(x)$. 
Each bitstring encodes the occupations of the spin-orbitals and is identified with a Slater determinant $|x\rangle$. 
The collection of sampled determinants defines a subspace $\mathcal{S}$ of the full Hilbert space, and projecting the Hamiltonian as $P_{\mathcal{S}} H P_{\mathcal{S}}$ allows one to obtain an approximate ground-state vector by diagonalising this reduced operator.

Measurements of the state $|\tilde{\Psi}\rangle$ inevitably contain errors, and near-term quantum devices often produce bitstrings that violate conserved quantities of the electronic problem, such as particle number or spin. 
Since the ground state of $H$ resides in a fixed symmetry sector, SQD introduces a configuration-recovery step that maps the raw samples to determinants consistent with these symmetries. 
Bitstrings lying outside the target sector are probabilistically corrected using a transformation informed by the current estimate of the orbital-occupation vector. 
The outcome is a set of recovered determinants with the correct particle number and spin.

This recovery step appears within an iterative loop. 
Each iteration begins with a fresh collection of quantum-generated bitstrings, applies the symmetry-based correction procedure to obtain a recovered ensemble $X_R$, and uses this ensemble to define the subspace $\mathcal{S}$. 
After diagonalising the projected Hamiltonian $P_{\mathcal{S}} H P_{\mathcal{S}}$ and updating the orbital-occupation vector implied by the resulting approximate eigenvector, the process repeats. 
Iteration gradually aligns both the sampling distribution and the recovered determinant set with the structure of the true ground-state wave function.

The SQD workflow can be strengthened by introducing a batching procedure that augments the basic cycle of sampling, configuration recovery, and subspace diagonalisation. 
Rather than constructing a single subspace from the recovered ensemble $X_R$, the method draws several independent batches $S^{(k)}$ from $X_R$ and diagonalises the projected Hamiltonian within each of them. 
This modification allows multiple diagonalisations to be carried out in parallel and yields several approximate eigenstates per cycle, whose orbital-occupation vectors can be aggregated to produce a more stable update for the subsequent recovery step~\cite{robledo-moreno_chemistry_2025}.

The SQD workflow also includes a carryover mechanism: determinants whose projected amplitudes exceed a prescribed threshold in one iteration are automatically retained in the batch used for the next iteration. 
This ensures that configurations known to have appreciable weight are not lost during the sampling and recovery steps and provides a stable backbone for the iterative refinement of the occupation vector.

\section{Neural-network classifier implementation \label{sec:classifier-implementation}}

Here we summarise the implementation details of the neural-network classifier used in both cHCI and cSQD. 
The classifier is designed to be simple, lightweight, and retrained at every iteration so that it adapts to the evolving structure of the approximate wave function.


\textit{Input representation}. Each configuration is represented as a binary vector encoding the occupations of the molecular spin orbitals. 
Alpha and beta spin channels are stored either by interlacing or concatenating their occupation bits; this choice has no effect on performance, as all architectures considered in this work are insensitive to orbital ordering. 
The molecular orbitals are obtained from restricted Hartree--Fock calculations performed with PySCF. 
No additional preprocessing, normalisation, or feature engineering is applied to the input vectors.

\textit{Network architecture}. For all results reported in this work, we employ a compact MLP with the following structure:
\begin{itemize}
    \item Two fully connected layers of width 128 with ReLU activations.
    \item A dropout layer with rate 0.1 applied after each hidden layer.
    \item A final dense layer with two outputs followed by a softmax, producing class probabilities for the ``important'' and ``unimportant'' labels.
\end{itemize}

This architecture is computationally inexpensive while providing sufficient expressive power for the classification task.

\textit{Training objective and metrics}. The classifier is trained using a weighted cross-entropy loss. 
The weight assigned to each class is the inverse of its frequency in the labelled dataset, compensating for the pronounced imbalance between important and unimportant configurations.

In addition to the training loss, we monitor the precision--recall area under the curve (PR-AUC) for the positive class. 
This metric emphasises recall of important configurations, which is essential for maintaining the accuracy of the underlying variational method.

\textit{Data generation and training procedure}. At each iteration of cHCI or cSQD, the underlying algorithm produces a pool of candidate configurations. 
A random subset of this pool is merged with the current variational space, and the Hamiltonian projected onto this temporary space is diagonalised. 
CI amplitudes from this diagonalisation yield labels for the configurations in the random subset. 
The labels are assigned by enforcing fixed fractions of positive and negative examples within the training subset.

The classifier is trained from scratch at every iteration using this labelled data. 
A validation set comprising one tenth of the labelled configurations is held out during training to monitor performance and prevent overfitting. 
Retraining at every iteration ensures that the classifier remains aligned with the current approximated wave function.

\textit{Classifier-guided expansion of the variational space}. After training, the classifier is applied to all remaining configurations in the candidate pool. 
The variational space for the next iteration is expanded by including:
\begin{itemize}
    \item All configurations labelled as important during the training-data generation step.
    \item A fixed number $M$ of configurations with the highest predicted probability of being important among those not in the training subset.
\end{itemize}

This hybrid mechanism ensures that ground-truth important configurations are always included, while the classifier provides additional guidance to focus computational effort on the most relevant sectors of the configuration space.

\textit{Optimiser, learning-rate schedule, and training parameters}. Training uses the AdamW optimiser with a learning-rate schedule consisting of a linear warmup followed by cosine decay. 
The learning rate increases linearly from zero to its peak value over the first 3000 optimisation steps, after which it follows a cosine decay for an additional 20{,}000 steps. 
The peak learning rate used for all systems in this work is $10^{-4}$. Each classifier is trained for 125 epochs per iteration.

These settings provide stable training dynamics and consistent classification performance across all systems studied.

\textit{Hyperparameters}. For reproducibility, \cref{tab:classifier-hparams} summarises the hyperparameters used in all classifier-enhanced calculations.

\begin{table}[h]
\centering
\begin{tabular}{l c}
\hline
\textbf{Hyperparameter} & \textbf{Value} \\
\hline
Input encoding & Binary occupation vector \\
Hidden layers & 2 \\
Hidden dimension & 128 \\
Activation function & ReLU \\
Dropout rate & 0.1 \\
Output layer & 2-unit softmax \\
Loss function & Weighted cross-entropy \\
Positive-class metric & PR-AUC \\
Optimiser & AdamW \\
Peak learning rate & $10^{-4}$ \\
Warmup steps & 3000 \\
Cosine-decay steps & 20000 \\
Epochs per iteration & 125 \\
Training/validation split & 0.9 / 0.1 \\
Selection rule (inference) & Top-$M$ highest probabilities \\
\hline
\end{tabular}
\caption{Hyperparameters used in the neural-network classifier.}
\label{tab:classifier-hparams}
\end{table}

\section{Comparison of NN architectures}

\begin{figure*}[htb]
	\begin{center}
	 \includegraphics[width=2\columnwidth]{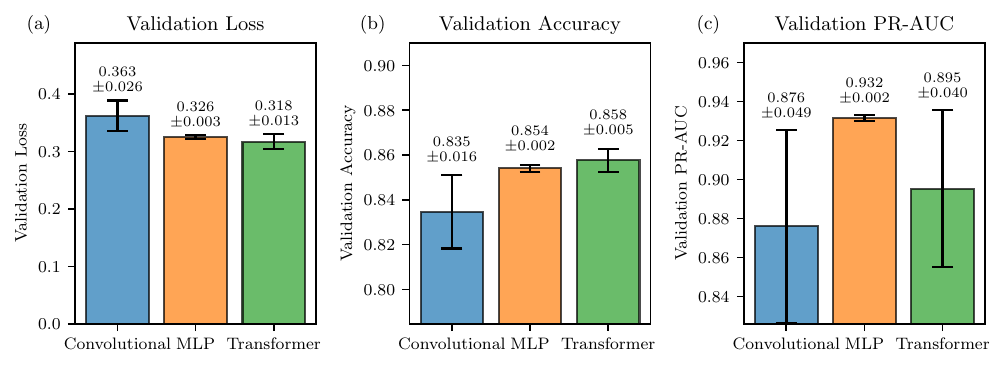}
	\caption{Comparison of neural-network architectures for Slater-determinant classification. 
    Three architectures are evaluated: a one-dimensional convolutional network (Convolutional), 
    a two-layer feed-forward network (MLP), and an attention-based model (Transformer). 
    Each architecture was trained 20 times with different random seeds; bars show the mean 
    and error bars indicate one standard deviation. 
    (a)~Validation loss. 
    (b)~Validation accuracy. 
    (c)~Precision--recall area under the curve (PR-AUC), which emphasises recall of important 
    configurations. 
    The MLP achieves the highest PR-AUC ($0.932 \pm 0.002$) and exhibits the lowest variability 
    across all metrics, making it the preferred architecture for the classifier used in cHCI and cSQD. 
    The convolutional network shows substantial run-to-run variation, consistent with the absence 
    of translational invariance in occupation-number vectors. 
    Data computed for N$_2$ at $R = \qty{2.1}{\angstrom}$ in the cc-pVDZ basis with $N_\mathrm{o} = 16$
    active spatial orbitals and $N_\mathrm{e} = 10$ active electrons.}
    \label{fig:3}
	\end{center}
\end{figure*}

We explored several neural architectures, including multilayer perceptrons (MLPs), one-dimensional convolutional networks, and transformer-based models, over a broad range of hyperparameters, see \cref{fig:3}. 
As expected, one-dimensional convolutions provided no advantage, since Slater-determinant occupation vectors possess no translational invariance. Transformer models likewise did not outperform simpler feed-forward networks.


\clearpage
\nocite{Varutti_in_preparation}
\bibliography{bibl}

\end{document}